\documentclass{kapproc} % Computer Modern font calls

\usepackage{procps} 

\usepackage[dvips]{graphicx}

\upperandlowercase

\normallatexbib %will produce bibliography entries as shown in the
\begin{document}

%------------ article title  ------------------->>

% If you use \\'s , please supply an alternate version of the title
% in square brackets, i.e., 
%\articletitle[Communism, Sparta, and Plato]
%{COMMUNISM, SPARTA,\\ and PLATO}

\articletitle[]{Disproportionation and Spin Ordering  \\
Tendencies in
N\lowercase{a}$_{\lowercase{x}}$C\lowercase{o}O$_2$
                      at $\lowercase{x}=\frac{1}{3}$}

%% optional, to supply a shorter version of the title for the running head:
%%\chaptitlerunninghead{}

%\author{J. Kune\v{s}, K.-W. Lee, and W. E. Pickett}

%% multiple authors may be separated with \\
%% \author{Samuel Bostaph\\
%% and Gregor Kariotis}

%------ author/affiliation choices -------------->>

%% Single author

% \author{}
%\affil{Department of Physics, University of California, Davis CA 95616}
% \email{}

%% Multiple authors, single affiliation

% \author{Samuel Bostaph}
% \author{George Lewis}
% \and   % <=== Type in \and before the last author so that `and' will
% \author{Cleon Jones}    % print between the last two authors
                           % in the table of contents.

% \affil{}

%% Multiple authors, multiple affiliations

% First method:
--------------

\author{J. Kune\v{s},$^{1,2}$ K.-W. Lee$^1$ and W. E. Pickett$^1$}
\affil{$^1$Department of Physics, University of California, Davis CA 95616}
\affil{$^2$Institute of Physics, Academy of Sciences,
   Cukrovarnick\'a 10, CZ-162 53 Prague, Czech Republic}

%\author{K.-W. Lee and W. E. Pickett}
%\affil{Department of Physics, University of California, Davis CA 95616
%      }

% Second method:
--------------

%\author{author\altaffilmark{1}, author\altaffilmark{2}, 
%         author\altaffilmark{1,3}}

%\altaffiltext{1}{Affiliation}
%\altaffiltext{2}{Affiliation}
%\altaffiltext{3}{Affiliation}

% Third method:
--------------

%\author{author\altaffilmark{1}, author\altaffilmark{2}, 
%         author\altaffilmark{1,3}}

%\affil{\altaffilmark{1}First affiliation, \ 
%\altaffilmark{2}Second affiliation, \
%\altaffilmark{3}Second affiliation}

%------ prologue, abstract, keywords ----------->>
% optional prologue
%\prologue{<text>}{<author, year>}

% optional abstract
\begin{abstract}
The strength and effect of Coulomb correlations in the
(superconducting when hydrated) $x\approx$1/3
regime of Na$_x$CoO$_2$ have been evaluated using the correlated band theory
LDA+U method.
Our results, neglecting quantum fluctuations, are:
(1) there is a critical $U_{c}$ = 3 eV, above which charge ordering
occurs at $x$=1/3, (2) in this charge-ordered state, antiferromagnetic
coupling is favored over ferromagnetic, while below $U_{c}$, ferromagnetism
is favored; and (3) carrier conduction behavior should be very
asymmetric for dopings away from $x$=1/3.  For $x < \frac{1}{3}$,
correlated hopping of parallel spin pairs is favored, suggesting a
triplet superconducting phase.  
\end{abstract}

% optional keywords
% \begin{keywords}
% Text, text...
% \end{keywords}

\section{Introduction}
Since the discovery of high temperature superconductivity in cuprates,
there has been intense interest in transition metal oxides with strongly
layered, (quasi) two-dimensional (2D) crystal structures and electronic
properties.  For several years now alkali-metal intercalated layered cobaltates,
particularly Na$_x$CoO$_2$ (NxCO) with $x\sim 0.50 - 0.75$, 
have been pursued for their thermoelectric properties.\cite{terasaki}
Li$_x$CoO$_2$ is of course of great interest and importance due to its
battery applications.
The recent discovery\cite{takada} and 
confirmation\cite{sakuraiornl,verify,chou}
of superconductivity in this system, for $x\approx$ 0.3
when intercalated with H$_2$O,
has heightened interest
in the NxCO system.  

The crystal structure\cite{ono,lynn,jorgensen} is based
on a 2D CoO$_2$ layer in which edge-sharing CoO$_6$ octahedra lead to a
triangular lattice of Co ions.
Na donates its electron to the CoO$_2$ layer, hence
$x$ controls the doping level of the layer: $x$=0 corresponds to Co$^{4+}$,
S=$\frac{1}{2}$ low spin ions with one minority $t_{2g}$ hole, and $x=1$
corresponds to non-magnetic Co$^{3+}$.  Nearly all reports of
non-stoichiometric materials quote values
of $x$ in the 0.3 - 0.75 range, and the materials seem generally to show
metallic conductivity.  
Reports of the magnetic behavior are of particular interest to us.
For $x$ in the 0.5 - 0.75 range, the susceptibility $\chi(T)$ is
Curie-Weiss-like (C-W) with reported moment of the order of magnitude
 1 $\mu_B$ per Co$^{4+}$\cite{takada,sakuraiornl} which
indicates the presence of correlated electron behavior on the Co ions.
Magnetic ordering at 22 K with very small ordered moment
has been reported for $x$=0.75\cite{x75order}  and
Wang {\it et al.} measured field dependence\cite{verify} that indicated 
the spin entropy of the magnetic Co system is responsible for the
unusual thermoelectric behavior.  Thus for $x\geq 0.5$ magnetic Co ions
and magnetic ordering give evidence of correlated electron behavior.  

However, for H$_2$O intercalated samples with $x \approx $0.3, 
({\it i.e.} the superconducting phase)
C-W behavior of $\chi$ vanishes.
\cite{sakuraiornl,chou,comment,kobayashi}.  
It is
extremely curious that the appearance of superconductivity correlates with
the disappearance of Co moments in the samples.  
From a single-band strongly interacting viewpoint, the
$x=0$ system corresponds to the half-filled triangular
lattice that has been studied extensively for local singlet (resonating
valence bond) behavior.\cite{moessner}   
The $x\approx 0.3$ region of 
superconductivity in NxCO is however well away from the half-filled
system, and the behavior in such systems is expected to 
vary strongly with doping level. 

There is
now a serious need to understand the electronic structure of the normal
state of the unhydrated material, and its dependence on the 
doping level $x$.  The electronic structure
of the $x$=1/2 ordered compound in the local density approximation (LDA)
has been described by Singh.\cite{singh0003}  Within LDA all Co ions are
identical (``Co$^{3.5+}$''), the Co $t_{2g}$ states are crystal-field split 
(by 2.5 eV) from the $e_g$
states, and the $t_{2g}$ bands are partially filled, 
consequently the system is metallic
consistent with the observed conductivity.  The $t_{2g}$ band complex is
$W \approx$ 1.6 eV wide, and is separated 
from the 5 eV wide O 2$p$ bands that lie just
below the Co $d$ bands.  Singh noted that the expected on-site Coulomb
repulsion $U$=5-8 eV on Co gives $U >> W$ and correlation effects can be
anticipated.  A value of $U\approx$ 4 eV was assumed by Wang, Lee, and 
Lee\cite{WLL} 
to justify a strongly correlated $t-J$ model treatment of this system.
While it must be kept in mind that the study of this system is still in its
infancy and no clear experimental data plus theoretical interpretation  
agreement has established the degree of correlation, we also take the 
viewpoint here that effects of on-site repulsion need to be assessed.

Although the experimental evidence indicates
nonmagnetic Co ion in the superconducting
material, most theoretical approaches\cite{WLL,thy1,thy2}  consider
the strongly interacting limit where not only is $U$ important,
it is large enough to prohibit double occupancy, justifying
the single band $t-J$ model.  Another question to address is whether the single
band scenario is realistic: indeed the rhombohedral symmetry 
of the Co site splits
the $t_{2g}$ states into $a_g$ and $e^{\prime}_{g}$ representations, but
the near-octahedral local symmetry leaves their band centers and widths
very similar.

In this paper we begin to address the correlation question using
the correlated band theory LDA+U method.  We focus on the $x \approx$1/3
regime where superconductivity emerges.  We find that $U \geq U_c =$ 3 eV 
leads to charge ordering at $x$=1/3 accompanied by antiferromagnetic
(AFM) spin order; of course, the fluctuations neglected
in the LDA+U method, the availability of three distinct sublattices for
ordering, and the tendency of the Na ions to 
order\cite{jorgensen} (which can mask
other forms of ordering at the same wavevector), can account for the
lack of ordering (or of its observation).  

\section{Method of Calculation}
Two all-electron full-potential electronic methods have been used.
The full-potential linearized augmented-plane-waves (FLAPW) 
as implemented in Wien2k code \cite{wien2k}
and its LDA+U \cite{nov01shick} extension were used. 
Valence and conduction $s, p$, and $d$ states were treated using
the APW+lo scheme \cite{sjo00}, while the standard LAPW 
expansion was used for higher $l$'s.
Local orbitals were added to describe Co 3$d$ and O 2$s$ and 
2$p$ states. 
The basis size
was determined by $R_{mt}K_{max}=7.0$. 
%Spin-orbit coupling as well as
%the additional LDA+U potential were included selfconsistently 
%in the second variational 
%step.\cite{shick}
%The cut-off energy for the second variation
%was set approximately 2.3 Ry above the edge of the valence band. 
In addition, the full-potential nonorthogonal local-orbital minimum-basis 
scheme (FPLO) of Koepernik and Eschrig\cite{koepernik99eschrig89} was also used
extensively.  Valence orbitals included Na $2s2p3s3p3d$, 
Co $3s3p4s4p3d$, and O $2s2p3s3p3d$.
The spatial extension of the basis
orbitals, controlled by a confining potential \cite{koepernik99eschrig89}
$(r/r_0)^4$, was optimized for the paramagnetic band structure and held
fixed for the magnetic calculations.
The Brillouin zone was sampled with regular mesh 
containing 50 irreducible k-points.
Both popular forms\cite{ani93czyzyk} of the LDA+U functional have been used to
assess possible sensitivity to the choice of functional, but in these studies
the differences were small. 
We do not consider interlayer coupling in the work presented here, which
allows us to use a single layer cell in the calculations.

\section{Results of Self-Consistent Calculations}

{\it LDA electronic structure at $x=\frac{1}{3}$.} 
The crystal field splitting of 2.5 eV puts the (unoccupied) $e_g$ states
(1 eV wide) well out of consideration.
The trigonal symmetry of the Co site
splits the $t_{2g}$ states into one of $a_g$ symmetry 
%\begin{eqnarray}
%|a_g> = \frac{1}{\sqrt 3} \bigl(|xy>+|yz>+|zx>\bigr) 
%\end{eqnarray}
%in the local coordinate system,
and a doubly degenerate $e^{\prime}_{g}$ pair.
The $a_g$ band is 1.5 eV wide (corresponding to
$t=0.17$ eV in a single band picture) while the
$e^{\prime}_{g}$ states have nearly the same band center but are only 1.3 eV
wide; hence they lie {\it within} the $a_g$ band.
As might be anticipated from the local octahedral environment, there is
mixing of the $a_g$ and $e_g^{\prime}$ bands throughout most of
the zone, and the $a_g$ DOS does not resemble that of an isolated band in
a hexagonal lattice. 
For the paramagnetic case $x=\frac{1}{3}$ corresponds to $\frac{8}{9}$
filling of the $t_{2g}$ band complex, resulting in hole doping into
the $e_g^{\prime}$ states as well as in the $a_g$ states. 

Singh found that ferromagnetic (FM) phases
seemed to be energetically favored for all noninteger 
$x$~\cite{singh0003}.  No ferromagnetism is seen in
this system, so NxCO becomes another member in small but growing list of
compounds\cite{list}  whose tendency toward FM is {\it overestimated} by LDA
because they are near a magnetic quantum critical point.
We confirm these FM tendencies within LDA for $x$=1/3, obtaining a 
half metallic FM state with a moment of 
$\frac{2}{3}\mu_B$/Co that is
distributed almost evenly on the three Co ions, which occupy two 
crystallographically distinct sites because of the Na position.  
The exchange splitting
of the $t_{2g}$ states 
is 1.5 eV, and the Fermi level (E$_F$) lies within the metallic minority
bands and just above the top of the fully
occupied majority bands.   
The FM energy gain is about 45 meV/Co.  With the majority bands filled,
the filling of the minority $t_{2g}$ bands becomes $\frac{2}{3}$, leading to
larger $e_g^{\prime}$ hole occupation than for the paramagnetic phase. 
We conclude that, in opposition to much of the theoretical speculation so far,
$x=\frac{1}{3}$ is necessarily a multiband ($a_g$ + $e_{g}^{\prime}$) system.
Our attempts using LDA to obtain self-consistent charge-ordered states, or 
AFM spin ordering, always converged
to the FM or nonmagnetic solution.

\begin{figure}[tbp]
%\vspace{0.2cm}
\begin{center}
\rotatebox{-90}{\resizebox{6cm}{6cm}{\includegraphics{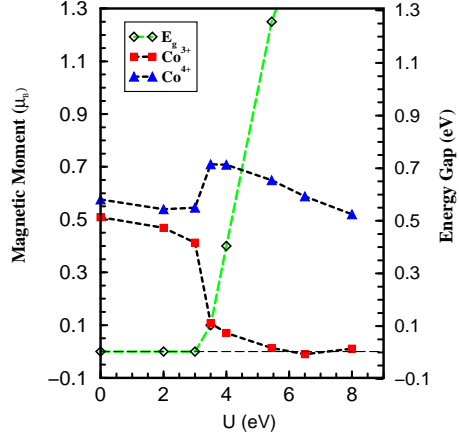}}}
\caption{Effect of the intraatomic repulsion $U$ on the magnetic 
moment of the Co1 and Co2 ions for ferromagnetic order.  The LDA+U
method in the FPLO code was used.  Disproportionation to formal
charge states Co$^{3+}$ and Co$^{4+}$ states occurs above $U_c$ = 3 eV.
}
\label{Moments}
\end{center}
\end{figure}

\begin{figure}[tbp]
%\vspace{0.2cm}
%\epsfxsize=7.5cm\centerline{\epsffile{strut3.eps}}
\flushleft
\rotatebox{-90}{\resizebox{2.7cm}{3.8cm}{\includegraphics{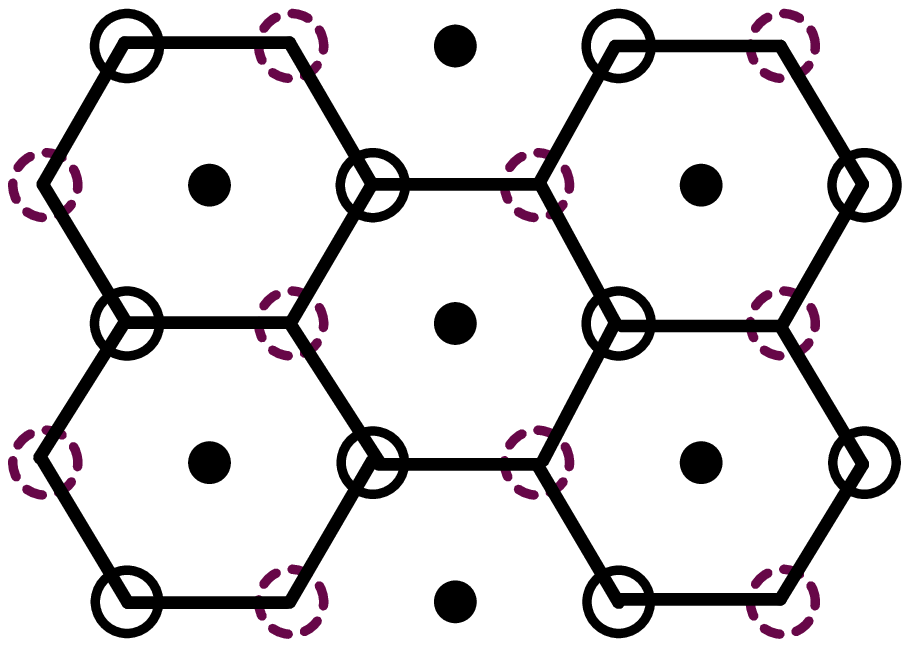}}}
\flushright
\vskip -30mm
\rotatebox{-90}{\resizebox{2.7cm}{3.8cm}{\includegraphics{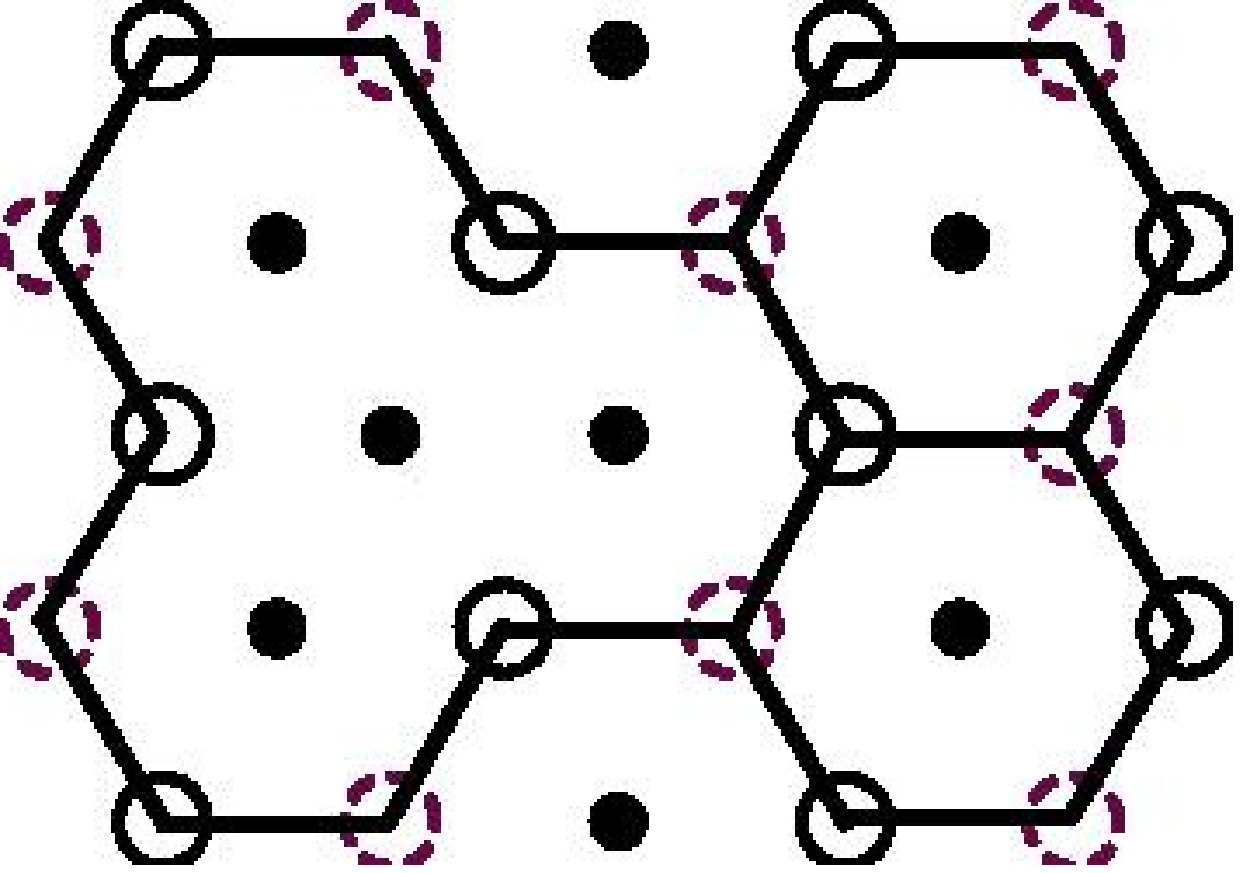}}}
\vskip 6mm
\flushleft
\rotatebox{-90}{\resizebox{2.7cm}{3.8cm}{\includegraphics{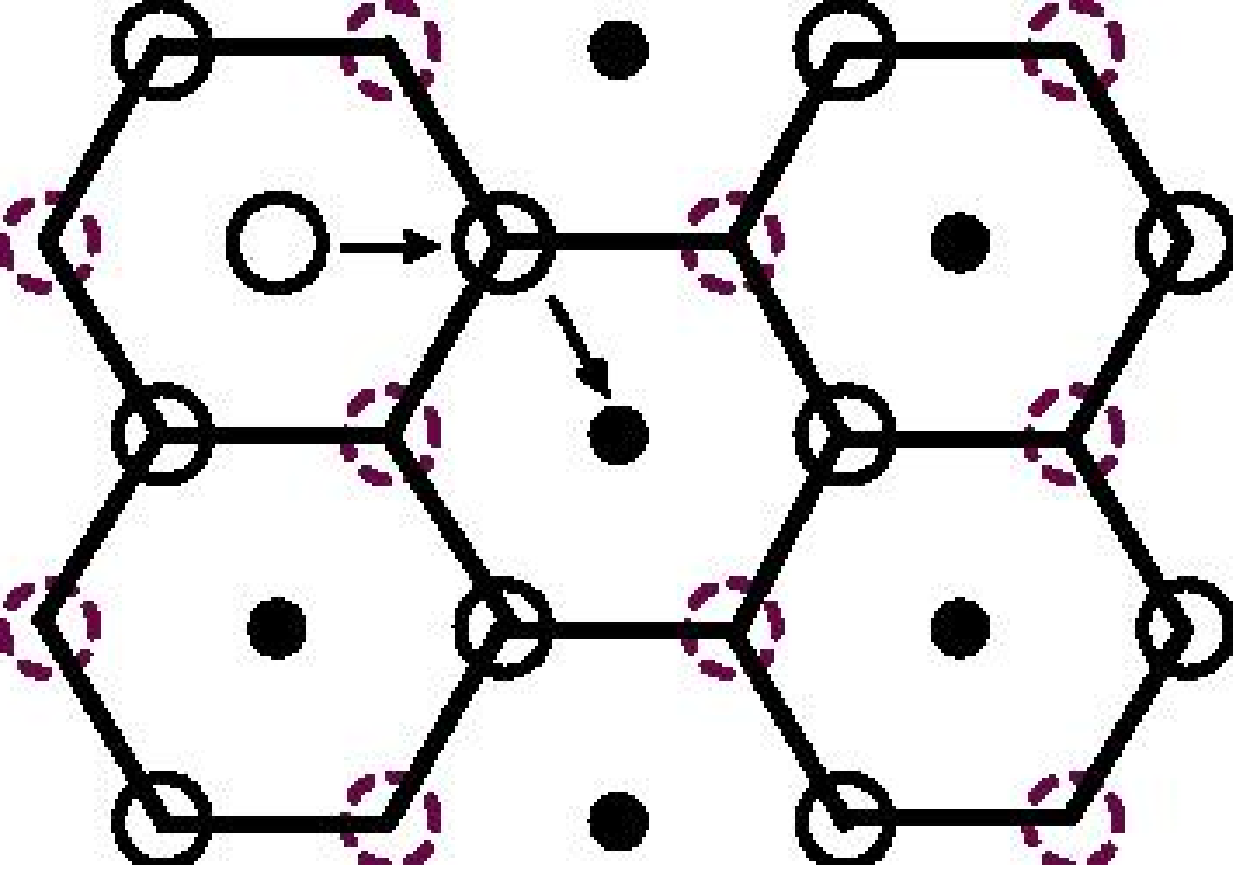}}}
\flushright
\vskip -30mm
\rotatebox{-90}{\resizebox{2.7cm}{3.8cm}{\includegraphics{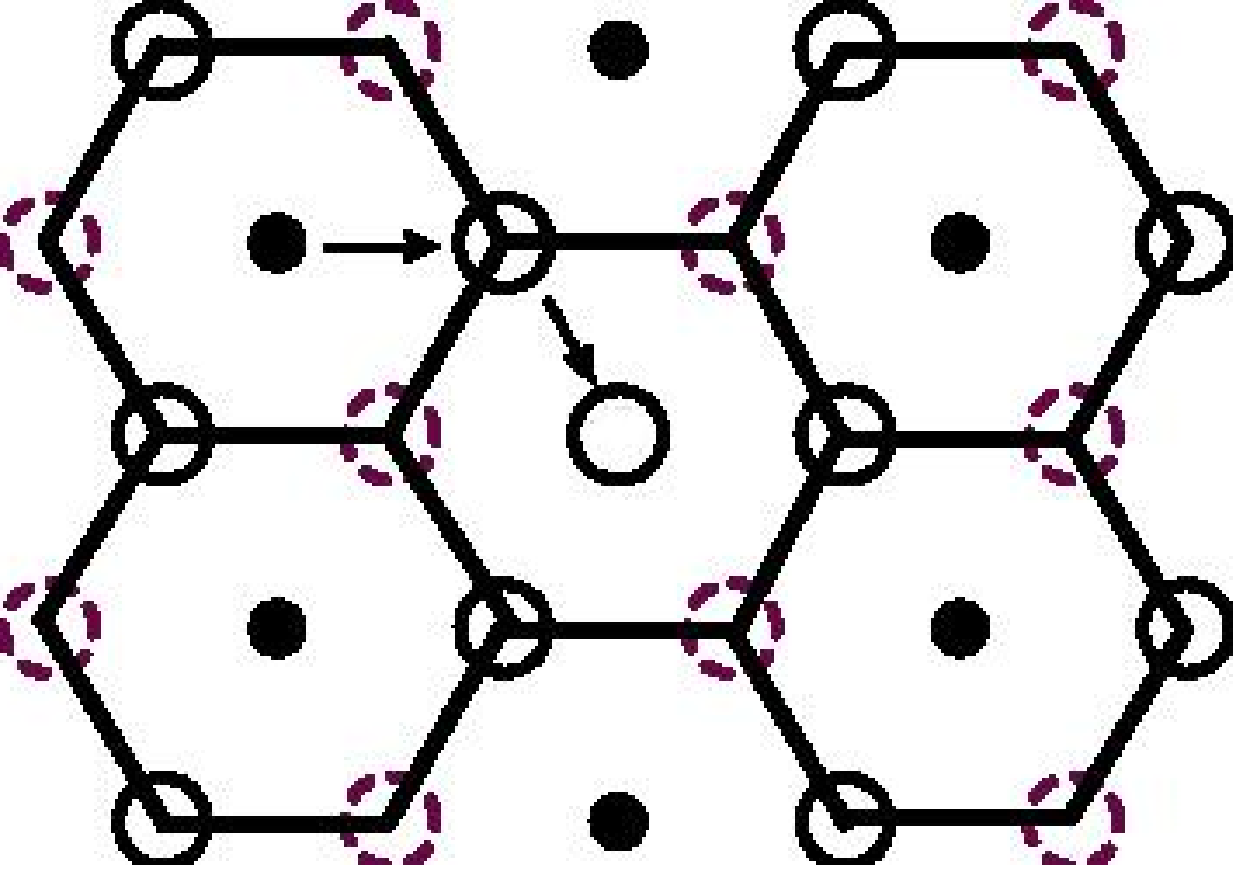}}}
\caption{  <Upper left> 
The charge ordered triangular Co lattice $\rightarrow$
honeycomb lattice, with antiferromagnetic
spin order designated by solid circles ($\uparrow$), dashed circles
($\downarrow$), and filled circles (Co$^{3+} S=0$ sites).  Lines denote 
nn magnetic couplings. <Upper right> addition of a $\uparrow$ electron
converts a Co$^{4+} S=\frac{1}{2}$ site to a nonmagnetic site.  Hopping 
of a neighboring hole to this site costs 4J in energy.
The lower two panels illustrate the correlated pair hopping process
after a $\uparrow$ hole is added to the system: hopping of the
hole to a neighboring site, followed by refilling of the site by the 
added hole, results in an identical (but translated) state.
}
\label{CO}
\end{figure}

{\it LDA+U Magnetic Structure and energies} 
First the behavior of the 
LDA+U results versus $U$ were studied (on-site exchange was kept fixed 
at 1 eV).  The dependence of the magnetic moment on $U$ (obtained from the
FPLO code) for FM ordering is shown in Fig. \ref{Moments}.  Recall that the
ordered array of Na ions in our cell gives two crystallographically
inequivalent Co sites.  For
$U < U_c \approx$ 3 eV, the moments are nearly equal and similar to   
LDA values (which is the $U\rightarrow$0 limit).  Above $U_c$, charge ordering 
accompanied by a metal-insulator (Mott) transition
occurs by disproportionation into nonmagnetic
Co$^{3+}$ and two $S = \frac{1}{2}$ Co$^{4+}$ ions.
Nonmagnetic Co$^{3+}$ states lie at the bottom of the
1 eV wide gap, with the occupied Co$^{4+}$ 
$e_{g}^{\prime}$ states 1-2 eV lower. 
After adding the on-site correlation to LSDA results, 
the hole on the Co$^{4+}$ 
ion occupies the $a_g$ 
orbital.  A possibility that we have not attempted would be to obtain
a solution in which the
hole occupies an $e_{g}^{\prime}$ orbital, in which case one should then
investigate orbital ordering
in addition to charge- and spin-ordering.

Reasonable estimates for cobaltates put $U$ at 5 eV or above,
so we now concentrate on results for $U$=5.4~eV, 
which we expect is near the
lower end of reasonable values.  For this value of $U$ both FM and
AFM spin-ordered solutions are readily obtained, with AFM energy 
1.2 mRy/Co lower than for FM order.
In terms of nn coupling on the charge-ordered honeycomb lattice, 
the FM - AFM energy difference corresponds
to $J$ = 11 meV.  Referring to the paramagnetic bandwidth identified 
above, the corresponding superexchange constant would be 4$t^2/U$
$\sim$ 20 meV.

{\it Discussion and Comparison with Experiment} 
These calculations establish that at $x$=1/3, there is a 
strong tendency to charge-order, and that there is a nn $J$ of 
antiferromagnetic sign; hence we consider as reference the 
$\sqrt 3 \times \sqrt 3$ charge-ordered  AFM state
shown in Fig. \ref{CO}.  
Considering the charge-ordering energies as 
dominant over the magnetic energies,
the fundamental problem at $x = \frac{1}{3}$ 
becomes the spin behavior of the half-filled honeycomb
lattice.  
Spin correlations and quantum fluctuations on the
honeycomb lattice have been considered by Moessner 
{\it et al.}\cite{qdimer} based on the quantum dimer 
model, where singlet-pairing regimes indeed are found 
in which the rms
magnetization on a site is strongly reduced.   Such pairing would
strongly suppress the Curie-Weiss susceptibility.

The foregoing discussion neglects
(among other aspects) the metallic nature of NxCO.
Now we consider ``doping'' away from $x$=1/3.  Addition of an electron
(of, say, spin up) converts a Co$^{4+} \downarrow$ to a Co$^{3+}$, that is,
it destroys a spin down hole which also was a potential carrier (if charge
order were lost).  This frees up a site for hopping of a neighboring hole,
but the energy cost of doing so is $4J$ (loss of two favorable $J$ and
gain of two unfavorable $J$) and thus is strongly inhibited.
Now we turn to the
removal of an $\uparrow$ electron (addition of a hole) corresponding to 
superconducting region
$x < \frac{1}{3}$, 
which has quite a different
effect.  This type of doping converts an inert Co$^{3+}$ to a Co$^{4+}$
that is surrounded by six Co$^{4+}$ sites with alternating spins.
Single particle hopping is disallowed (strictly speaking, 
it costs $U$); however,
correlated {\it parallel-spin pair hopping} as shown in Fig. \ref{CO}
returns the system to an equivalent state and 
therefore requires no net energy.  
This process suggests a tendency toward
triplet pairing superconductivity in this regime, although our results are too
preliminary to permit serious conclusions.
Although some progress on such 
processes might be made analytically within a single band model, the
multiband nature of NxCO turns even the relatively simple Hubbard model
on a triangular or honeycomb lattice into a formidable numerical
problem.

\section{Summary and Acknowledgments}
Now we summarize.  We have used the LDA+U method to evaluate the 
effects of Hubbard-like interactions in NxCO, and find charge
disproportionation and a Mott insulating state at $x=\frac{1}{3}$ when
fluctuations are neglected.  Nearest neighbor coupling $J \approx$  11 meV
provides AFM correlations.  Indications based on this ``mean field''
AFM charge-ordered state are for very different behavior for electron
or for hole doping relative to $x=\frac{1}{3}$; hole doping from this point
tends to favor parallel-spin pair-hopping and thus possible triplet
superconductivity.  Fluctuation effects may however be substantial.

We acknowledge important communications with R. T. Scalettar,
R. R. P. Singh, R. Cava,
B. C. Sales, and D. Mandrus.
J. K. was supported by National Science Foundation Grant 
DMR-0114818.  K.-W. L. and W. E. P. were supported by 
DOE Grant DE-FG03-01ER45876.

%% optional
%\section{Summary}

%% optional
%\begin{acknowledgments}
%...
%\end{acknowledgments}

%% appendix optional
%\appendix{This is the Appendix Title}
%This is an appendix with a title.

%\appendix{}
%This is an appendix without a title.

%
% Bibliography made with BibTeX:
%% kapalike is preferred if you have used \kluwerbib, above.
%% Otherwise you may use any .bst style your editor approves.

%This will allow many Bib\TeX\ bibliographies in one book.
%See the documentation, edbk.doc, for more information.

%\bibliographystyle{kapalike}
%\chapbblname{<name of .bbl file>}
%\chapbibliography{<name of .bib file>}

%or 
%\begin{chapthebibliography}{<widest bib entry>}

%\end{chapthebibliography}

\end{document}